\documentclass[aps,prl,twocolumn,showpacs,superscriptaddress,groupedaddress]{revtex4-1}
\usepackage{graphicx}
\usepackage{amssymb}
\usepackage{amsmath}
\usepackage{latexsym}
\usepackage{ulem}
\usepackage{color}
\usepackage{dcolumn}
\usepackage{subfigure}
\usepackage[T1]{fontenc}
\usepackage[utf8]{inputenc}
\usepackage[english,frenchb]{babel}
\usepackage[colorlinks=true,linkcolor=blue,citecolor=blue]{hyperref}%
\usepackage[toc,page]{appendix}

\usepackage{bm}        
\usepackage{amssymb}   
\begin{document}

\title{RKKY couplings in the Lieb lattice: flat-band induced frustration}

\author{G. Bouzerar}
\email[E-mail:]{georges.bouzerar@neel.cnrs.fr}
\affiliation{CNRS and Université Grenoble Alpes, Institut NEEL, F-38042 Grenoble, France}        
\date{\today}
\selectlanguage{english}
\begin{abstract}
Despite their dispersion-less character, flat bands (FBs) are often at the heart of remarkable physical phenomena. Because, FBs may be responsible for unconventional quantum electronic transport, our purpose is to investigate their impact on the RKKY couplings ($J(\bf{R})$). As a good candidate, we choose the The Lieb lattice which consists of one four-fold coordinated atom (A) and two two-fold coordinated atoms (B,C) per unit cell. As in graphene, at the neutrality point, $J(\bf{R})$ is found to fall off as $1/R^3$. The coupling between a pair of impurities on sublattice A are found ferromagnetic and isotropic. In contrast, for impurities on the two-fold coordinated sites, $J(\bf{R})$ reveals (i) a strong angular anisotropy and (ii) can be both ferromagnetic (F) or antiferromagnetic (AF). The AF character introduced by the FB can largely dominate, and introduces frustration effects. The FB is found to have a drastic impact on $J(\bf{R})$ for impurities located on (B,C) sublattices. We have also addressed the effect of tuning the carrier density that reveals remarkable features as well.
\end{abstract}
\pacs{75.50.Pp, 75.10.-b, 75.30.-m}
\maketitle

\section{Introduction}

Over the past decade, systems which possess flat or dispersionless bands in their electronic band structure have gained a considerable attention. 
This novel family of emerging systems do not cease to challenge and fascinate the condensed matter physics community. 
In these systems, the flatness of the bands results from destructive quantum interferences that lead to a vanishing electron group velocity, and thus, to the quenching of the kinetic energy. Flat bands are at the heart of a plethora of remarkable and intriguing quantum physical phenomena.
First, flat band systems may host an unconventional type of superconductivity that could pave a unique route towards high-$T_C$ materials \cite{julku,peri,peotta,lizardo,miyahara,aoki}. For instance, this unusual and unexpected form of superconductivity has been reported in twisted bilayer and trilayer graphene near magic angles \cite{cao,chen1,yankowitz,park}. Near these particular twisting angles, the electronic band structure exhibits nearly flat bands in the vicinity of the Fermi energy. The superconductivity associated to flat bands has been shown to be of topological (or geometrical) nature and presents a singular linear dependence of the order parameter (or of the critical temperature) with the strength of the attractive electron-electron interaction parameter \cite{julku,peri,peotta}.
Flat bands are also responsible for the appearance of magnetic phase in the presence of arbitrary small values of the electron-electron interaction strength.
This is the so called and well known flat-band induced ferromagnetism. It has been originally predicted by Lieb \cite{lieb} and later, it was put into a more broad and general context by Tasaki and Mielke \cite{tasaki,mielke}. Within the numerical renormalization group approach, accurate for strongly correlated systems, the issue of flat band ferromagnetism was theoretically addressed in the case of fermions in the Lieb lattice \cite{noda}. Besides the CuO$_2$ planes in cuprates, so far, there is no true two dimensional Lieb lattice material, but it is worth mentioning that it has been recently reported that the Lieb lattice could be experimentally designed, realized, in the framework of covalent-organic compounds \cite{jiang}. Flat bands, also explain the origin of the unusual ferromagnetism observed in a wide family of non magnetic systems and also known as $d^0$ ferromagnetism \cite{venkatesan,young1,makarova,bouzerard0}. Among the other fascinating physical phenomena associated to flat bands, we can quote the existence of non trivial topological phases \cite{kang,gao,liu,tang,sun,pal,neupert}, the Wigner crystallization phenomenon \cite{wu1,chen} and even the possibility of super-metallic phases where one would expect insulating ones \cite{bouzerarSM12}. The wealth and the intriguing physics that take place in these exotic systems motivate the search for efficient procedures and strategies for flat-band material engineering \cite{hase,xu,mizoguchi,vergel,kerelsky}.

\begin{figure}[t]\centerline
{\includegraphics[width=1.0\columnwidth,angle=0]{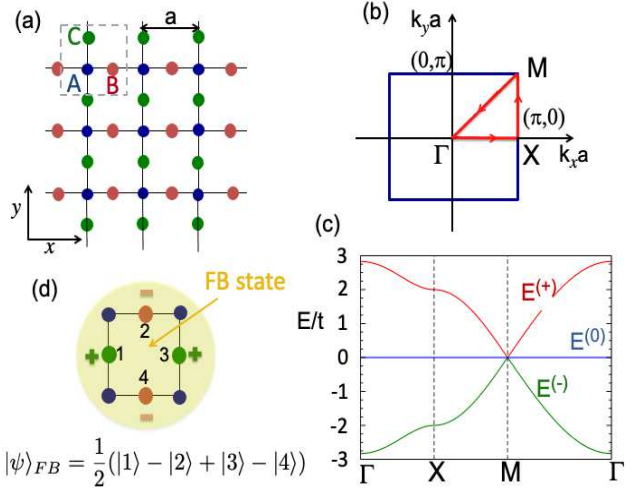}}
\caption{(Color online) (a) An illustration of the Lieb lattice that consists of a square lattice with 3 atoms/unit cell (A,B and C). The first Brillouin zone (BZ) is depicted in (b).The dispersion of the bands, $E^{-}$ (valence band), $E^{+}$ (conduction band), and $E^{0}$ (flat-band) along the BZ path $\Gamma$$\rightarrow$X$\rightarrow$M$\rightarrow$$\Gamma$ is represented in (c). A typical flat-band eigenstate (plaquette state) is illustrated in (d), it has non vanishing weight on 4 sites only, labelled from 1 to 4.
}
\label{fig1}
\end{figure} 
Because, flat bands conceals many exotic properties and hosts counter intuitive phenomena, our purpose is to address the nature of the RKKY couplings \cite{ruderman,kasuya,yosida} in a system that has a flat band (FB) in its electronic band structure. More precisely, one of the main goals is to understand how the couplings are affected by the presence of the FB and another is to evaluate quantitatively the contribution to the couplings that originates from the off-diagonal matrix elements (inter-band) that involve the FB.
A good candidate to investigate such an issue is the Lieb lattice. The Lieb lattice consists of 3 atoms per unit cell, one is four fold coordinated (A site) and the other two 
have only two nearest neighbors as it is illustrated in Fig.~\ref{fig1}. The Lieb lattice naturally possess a FB at $E^0({\bf k})=0$ and two symmetric dispersive
bands $E^+({\bf k})$ and $E^-({\bf k})$, with $E^+({\bf k})= -E^-({\bf k})$ that result from particle-hole symmetry. It is worth noticing that the Lieb lattice can be seen as a 25\% depleted square lattice, where only A-type sites are removed.
Recently, it has been reported that the electronic transport properties at the FB energy in various systems, including the Lieb lattice, are of unconventional type. It was shown, that the FB is responsible for the existence of a supermetallic phase (with a vanishing Drude weight), that is both robust against (i) disorder and (ii) the presence of a gap between the FB and the dispersive bands \cite{bouzerarSM12}. Thus, we believe that the FB should also have a remarkable impact on the nature of the magnetic couplings.

Electrons on the Lieb lattice are modeled by a nearest neighbor tight binding Hamiltonian that reads,
\begin{eqnarray}
\widehat{H}=-t \sum_{\left\langle ij\right\rangle,s} c_{is}^{\dagger}c^{}_{js} +h.c.,
\label{hamilt}
\end{eqnarray}
$t$ is the hopping integral, $\left\langle ij\right\rangle$ denotes nearest neighbor pairs ((A,B) and (A,C)),
 c$_{i\sigma}^{\dagger}$ creates an electron with spin $\sigma$ at site \textbf{R}$_{i}$. Here, we consider the absence of A-A, B-B or C-C hoppings (next nearest neighbor hopping) in order to focus on the effect of a true flat band. In what follows, we will also ignore the spin label, irrelevant here, and consider spinless fermions.

\begin{figure}[t]\centerline
{\includegraphics[width=0.95\columnwidth,angle=0]{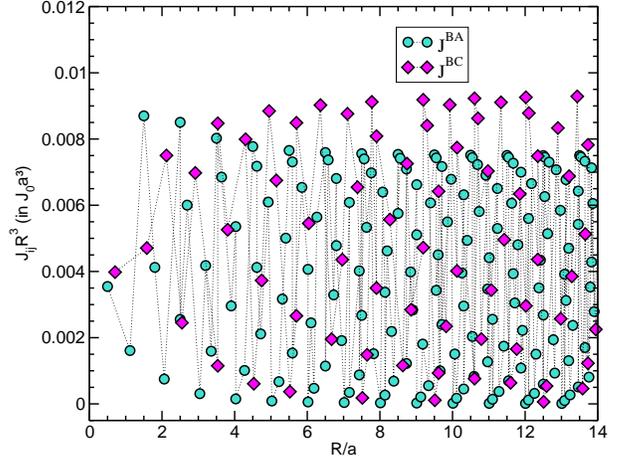}}
\caption{(Color online) $J_{ij}R^3$  as a function of R/a, the distance between pairs of impurities located on different sublattices. $J_{ij}$ are the  couplings between the (B,A) and (B,C) pairs. Positive (resp. negative) couplings correspond to anti-ferromagnetic (resp. ferromagnetic) exchanges. The chemical potential is set to $\mu=0$ and the couplings are expressed in unit of $J_{0} = \frac{J^2}{t}$.}
\label{fig2}
\end{figure}

\section{The RKKY couplings in the Lieb lattice}

The RKKY coupling between two impurities located respectively at site $R_{\bf i}$ of sub-lattice $\alpha$ and $R_{\bf j}$ of sub-lattice $\beta$ reads \cite{rkky-coupling},
\begin{eqnarray}
J_{\bf ij}^{\alpha\beta} =-\frac{J^2}{\pi} \int_{-\infty}^{\infty}  \Im \left[ G_{\bf ij}^{\alpha\beta}(\omega) G_{\bf ji}^{\beta\alpha}(\omega) \right]f(\omega)d\omega,
\label{coupling}
\end{eqnarray}
where $J$ denotes the local coupling between the spin of the magnetic impurity and that of the itinerant carrier. $f(\omega)=(e^\frac{(E-E_F)}{k_BT}+1)^{-1}$ is the Fermi distribution
 and the Green's function $G_{\bf ij}^{\alpha\beta}(\omega)= \langle {\bf i}, \alpha \vert ((\omega+i\eta)\hat{1}-\widehat{H})^{-1} \vert \beta {\bf, j}  \rangle$.
In this section, we consider the case $E_F=0$ and the temperature is set to $T=0\,K$.
We stress that $J_{\bf ij} \ge 0$ (resp.  $J_{\bf ij} \le 0$) means antiferromagnetic (resp. ferromagnetic) coupling. 

The straightforward diagonalization of the Hamiltonian given in eq. (\ref{hamilt}) leads to two dispersive bands $E^\pm({\bf k})=\pm 2t f_0({\bf k})$ where $f_0({\bf k})=\sqrt{\cos^2(k_xa/2)+\cos^2(k_ya/2)}$ and a flat-band with energy $E^0({\bf k})=0$. Their respective eigenstates are,
 $\vert \Psi ^\pm({\bf k})\rangle = \frac{1}{\sqrt{2}}  (\pm c_x(\textbf{k}) \vert B,{\bf k} \rangle \pm c_y(\textbf{k}) \vert C,{\bf k} \rangle  +  \vert A,{\bf k} \rangle )$ and
 $\vert \Psi ^0({\bf k})\rangle = -\frac{1}{\sqrt{2}}  (c_y(\textbf{k}) \vert B,{\bf k} \rangle - c_x(\textbf{k}) \vert C,{\bf k} \rangle ) $, where
 $ c_u({\bf k})) = \cos(k_ua/2)/ f_0({\bf k})$ ($u=x,y$) and $\vert \alpha ,{\bf k} \rangle =\frac{1}{\sqrt{N}} \sum_{\bf i}  e^{\bf k.R_i}  \vert \alpha ,{\bf i} \rangle, $ ($\alpha$ = A, B and C), $N$ being the number of unit cells.
 
\begin{figure}[t]\centerline
{\includegraphics[width=0.95\columnwidth,angle=0]{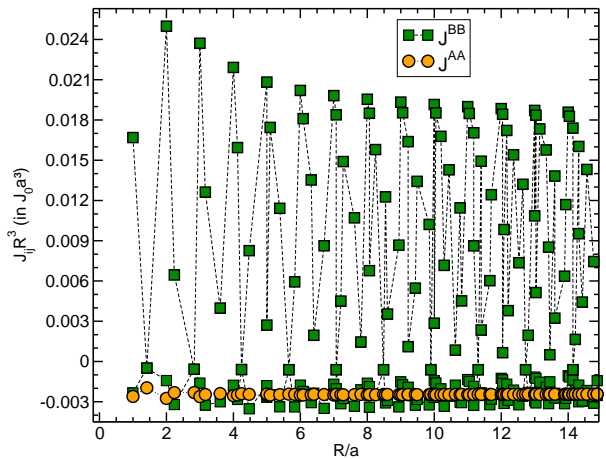}}
\caption{(Color online) 
$J_{ij}R^3$ as a function of R/a, the distance between pairs of impurities. Here, $J_{ij}$ are the couplings between the (A,A) and (B,B) pairs.
The couplings are expressed in unit of $J_{0} = \frac{J^2}{t}$.
}
\label{fig3}
\end{figure}

 In Fig.~\ref{fig2} and Fig.~\ref{fig3} are depicted the couplings as a function of the distance between the magnetic impurities as obtained from the exact calculation of the Green's functions. Notice that $J^{CA}$ and $J^{CC}$ are not shown, since there are respectively identical to $J^{BA}$ and $J^{BB}$.
First, we observe that both $J^{BA}$ and $J^{BC}$ exhibit large oscillations as we vary the distance but they keep the same sign, they are antiferromagnetic. In addition, the couplings appear to fall off as $1/R^3$.This is in contrast to the conventional $1/R^2$ behavior expected in 2D systems, but it agrees with what has been found in the case of pristine graphene at the Dirac point \cite{sherafati1,sherafati2,saremi,kogan}. In Fig.~\ref{fig3}, we observe significant differences between $J^{BB}$ and $J^{AA}$. As in previous cases, both couplings reveal a $1/R^3$ decay, as it will be confirmed analytically in the next section, but $J^{AA}$ is free from oscillation and stays ferromagnetic. More precisely, from a fit of the data for large distances, we find $J^{AA} \approx -2.5 \,10^{-3} \frac{a^3} {R^3} J_0$ where we have defined $J_0= \frac{J^2}{t} $. In contrast, $J^{BB}$ can be either ferromagnetic or antiferromagnetic. However, the antiferromagnetic couplings are found strongly
oscillating as the distance varies, they can reach much larger values, almost one order of magnitude, than the ferromagnetic ones.
This feature is in contrast with what has been reported in the case of graphene, where the couplings between impurities are always ferromagnetic (resp. antiferromagnetic) when they are located on the same (resp. different) sublattices \cite{sherafati1,sherafati2,saremi}. As it will be seen in the next section, the presence of the flat band is at the origin of this crucial difference.

\begin{figure}[t]\centerline
{\includegraphics[width=1.0\columnwidth,angle=0]{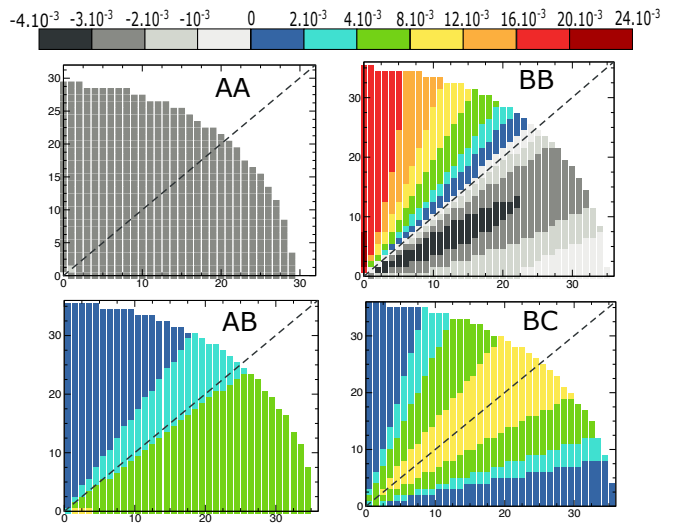}}
\caption{(Color online) 
2D color plot of $J_{ij}R^3$ for AA, BB, AB and BC pairs. The first impurity is located at the origin (A or B site) and the second impurity 
runs through the Lieb lattice. The horizontal (resp. vertical) axis correspond to the $x$ (resp. $y$) coordinate of the second impurity.
The color at the position of the second impurity provides the strength and sign of the coupling.The couplings are expressed in unit of $J_{0} = \frac{J^2}{t}$.}
\label{fig4}
\end{figure}

To have a better understanding of the nature of the magnetic couplings, we have realized in Fig.~\ref{fig4}, a two dimensional color plot. The horizontal and vertical axis correspond to the spatial coordinates of the second magnetic impurity, assuming that the first one is located at the origin. The color indicates the value and the sign (F or AF) of the coupling. We define $\theta$ the angle between ${\bf R}={\bf R_i}-{\bf R_j}$ and the horizontal axis ($x$-axis). In the the color plot we have restricted ourselves to $0 \le \theta \le \frac{\pi}{4}$, the couplings for the other angles can be deduced from the lattice symmetries.
A first glance at Fig.~\ref{fig4}, reveals that both $J^{BB}$ and $J^{AB}$ are asymmetric with respect to the $\theta=\frac{\pi}{4}$ line, as it could have been anticipated from the  Lieb lattice symmetry. Let us now discuss each panel individually. The (AA) panel reveals a uniform color that confirms what has been seen in Fig.~\ref{fig3}, $J^{AA}$ is ferromagnetic and fully isotropic. This is in contrast with the case of graphene where $J^{AA}$ depends on the factor $(1 + \cos(\bf{(K-K' )\cdot R}))$ where K and K' are the two inequivalent Dirac points in the  honeycomb lattice BZ \cite{sherafati1,sherafati2,saremi}. In graphene, the oscillations result from inter-valley matrix elements. In the Lieb lattice, the situation differs, the 3 bands cross each other at a single point, the M point ($M=(\frac{\pi}{a},\frac{\pi}{a})$) located at the corner of the Brillouin zone, as it is illustrated in Fig.\ref{fig1}.
We now discuss the case of the $J^{BB}$ couplings as they are depicted in the (BB) panel. For a fixed distance R, we clearly note a strong directional anisotropy of the couplings. Below the $(1,1)$ line (or $\theta=\frac{\pi}{4}$), the couplings are ferromagnetic while antiferromagnetic above this line. The ferromagnetic couplings reach their maximum at $\theta \approx \frac{\pi}{8}$, where $J^{BB} \approx - 3\,10^{-3} \frac{a^3} {R^3} J_0$.
 Above the (1,1) line, the coupling changes sign and rapidly increases as $\theta$ approaches $\frac{\pi}{2}$ ($y$-axis) where its maximum is reached. The amplitude of the variation of $J^{BB}\,R^3$ is only about $3\,10^{-3} J_0 \,a^3$ in the ferromagnetic region, while it is approximately  $24\,10^{-3} J_0 \,a^3$ for the antiferromagnetic couplings, in agreement with what has been observed in Fig.~\ref{fig3}. Below the (1,1) line, we notice that the maximum of $J^{BB}\,R^3$ is reached at $\theta \approx \frac{\pi}{8}$. Remark as well that the $J^{CC}$ couplings can be directly obtained from (BB)-couplings by applying a rotation around the $z$-axis (perpendicular to the Lieb lattice plane): $\theta \rightarrow  \frac{\pi}{2} -\theta$. Thus, in this case, the couplings are ferromagnetic above the $(1,1)$ line.
In the next panel of Fig.~\ref{fig4} ('AB'), it is found that the AF $J^{AB}$ couplings are also asymmetric with respect to $\theta=\pi/4$ and decrease gradually as $\theta$ increases. For a fixed distance R, the maximum is along the $x$-direction and the minimum along the $y$-direction. The amplitude of variation of $J^{AB}R^3$ is about $8\,10^{-3} J_0 a^3$. Again, $J^{AC}$ couplings can be deduced by a $\pi/2$ rotation. Regarding the AF $J^{BC}$ couplings, the situation is differs, we recover, as it could have been anticipated the lattice symmetry with respect to the (1,1) direction. The minimum of $J^{BC}R^3$ (for a given R) is reached for both $\theta = 0$ and $\frac{\pi}{2}$ and its maximum is along the (1,1) axis. In this case, the amplitude of variation of $J^{BC}R^3$ is found to be approximately $9 \,10^{-3} J_0 a^3$.

We summarize the main findings of this section. If we ignore for a moment the existence of a FB, by analogy with the case of graphene we would have expected that both $J^{BC}$ and $J^{BB}$ to be of ferromagnetic nature. In contrast, we find the opposite for $J^{BC}$ and even more intriguing results for $J^{BB}$, which (i) reveal a very strong angular anisotropy and (ii) both, a weak ferromagnetic couplings region and strong AF one. At this stage, it is interesting to discuss the role and impact of the presence of the flat-band at $E=0$.

\section{Analytical expression of the RKKY couplings and effects of the flat-band}

In this section, our aim is to clarify the origin of unexpected features found in Fig.~\ref{fig4}. For that purpose, we propose to analyze in details the different contributions to the couplings and derive the analytical expressions of the couplings as well. 

Because, the Fermi level is $E_F=0^+$ (results are identical for $E_F=0^-$), we are left with two different contributions to the exchange. One can write,
\begin{eqnarray}
J_{\bf ij}^{\alpha\beta} =-2J^2 \Big(  \sum_{\bf k k'} \frac {1}{\epsilon^0({\bf k} )} C_{\alpha\beta}^{0+} ({\bf k},{\bf k'},
{\bf R}) +\, \nonumber \\
\sum_{\bf k k'}  \frac {1}{\epsilon^0({\bf k}) +\epsilon^0({\bf k'} )}  C_{\alpha\beta}^{+-} ({\bf k},{\bf k'},
{\bf R})
\Big),
\label{coupling2}
\end{eqnarray}
where the matrix element $C_{\alpha\beta}^{\lambda\lambda'} ({\bf k},{\bf k'},
{\bf R})=\langle {\bf i} \alpha \vert  \Psi^\lambda_{\bf k}\rangle \langle  \Psi ^\lambda_{\bf k} \vert \beta {\bf j}  \rangle
\langle {\bf j} \beta \vert  \Psi^ {\lambda' }_{\bf k'}\rangle \langle  \Psi ^{\lambda'}_{\bf k'} \vert \alpha {\bf i}  \rangle$ with $\lambda, \lambda'= +, - $ or $0$,
where the label '-' corresponds to the valence band, '+'  to the conduction band and $'0'$ to the FB.
The second term in eq. (\ref{coupling2}) is the inter-dispersive band contribution and the first one is the flat-band contribution.
Notice also that  the matrix element $C_{{\bf i}\alpha,{\bf j}\beta}^{0+}$ is zero if at least one of the two impurities is located on a A-site. Thus, the contribution due to the flat-band affects only the (B,B) ,(C,C) and (B,C) pairs. The expressions of $C_{\alpha\beta}^{\lambda\lambda'}$ can be found in the appendix A.

In what follows, we define $J_{1}^{\alpha \beta} (\bf R)$ (resp.  $J_{2}^{\alpha \beta} (\bf R)$ ) as the first (resp. second) term in eq. (\ref{coupling2})
To allow the analytical calculations of the couplings, we start with the linearization of the dispersions in the vicinity of the neutrality point (M point in the BZ). The details are provided in the Appendix B.
For a pair of impurities located respectively on the $\alpha$ and $\beta$-sublattices, we write,
\begin{eqnarray}
J_{l}^{\alpha\beta} ( {\bf R})=- \frac{J_0}{8\pi^2}\frac{a^3}{R^3 }
 f_{l}^{\alpha\beta} (\theta),
\end{eqnarray}
where $l=1,2$ and we recall that $\theta$ is the angle between the $x$-axis and ${\bf R}$.

For a (A,A) pair, we have obtained $f_{1}^{AA} (\theta)=0$ (as expected) and $f_{2}^{AA} (\theta)= C$,  
and for a $(B,B)$ pair we have found $f_{1}^{BB} (\theta)=2\sin(\theta)\cos(2\theta)$ and $f_{2}^{BB} (\theta)= (\frac{\pi}{16} +4C)\cos^4(\theta)+2C(-3\cos^2(\theta)+1)$ where $C=1-\frac{\pi}{4}$. The analytical expression of $f_{l}^{\alpha\beta} (\theta)$ for the couplings for the other type of pairs is given in the appendix B. Let us discuss our findings.
\begin{figure}[t]\centerline
{\includegraphics[width=1.\columnwidth,angle=0]{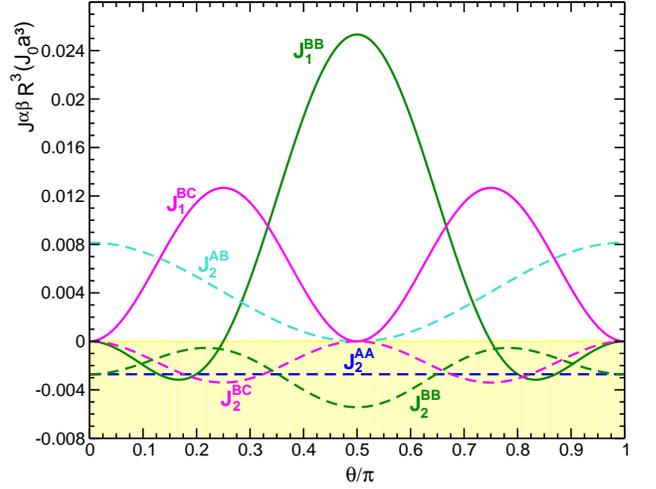}}
\caption{(Color online) 
$J^{\alpha\beta}R^3$ as a function of the angle $\theta$ between $ {\bf R = R_{i}-R_{j}}$  (${\bf R_{i}}$ and ${\bf R_{j}}$ are the impurity positions) and the horizontal axis ($x$-axis).
$J^{\alpha\beta}$ is the coupling between the impurities and $(\alpha,\beta)$ correspond to 4 different types of pairs (A,A), (B,B), (A,B) and (B,C).
$J_2$ couplings correspond to the inter dispersive bands contributions VB $\leftrightarrow$ CB and $J_1$ to FB $\leftrightarrow$ CB or FB $\leftrightarrow$ VB, where VB, CB and FB are respectively the valence band, the conduction band and the flat-band. The couplings are expressed in unit of $J_{0} = \frac{J^2}{t}$.
 }
\label{fig5}
\end{figure} 
First, in all cases the analytical calculations confirm the $1/R^3$ decay of the couplings. 
To ease the discussion, in Fig.~\ref{fig5}, is plotted $J_{1}R^3$ and $J_{2}R^3$, as a function of the angle $\theta$ for different kind of pairs $(\alpha,\beta)$. 
Again, in the case of a (A,A) pair the only contribution is the inter-dispersive band one. It is $\theta$-independent as found numerically before,
and the analytical expression gives $J^{AA}R^3 = -2.7 \,10^{-3}J_0 a^3$ which is very close to the value obtained from the full calculation (exact dispersions), $J^{AA} R^3 =-2.5 \,10^{-3}J_0 a^3$. In the case of $J^{AB}$ that also reduces to the $J_{2}$ contribution, the coupling is anti-ferromagnetic and has a clear $\theta$-dependence. Its reaches its maximum along the $x$-direction and vanishes along the $y$-axis.
On the other hand, as it is the case for $J^{AA}$, $J_{2}^{BB}$ is ferromagnetic which is what we would have expected in the absence of the FB, but it is strongly anisotropic as well. For a fixed R, this coupling is maximum (in amplitude) at $\theta=\pi/2$ with a value almost twice that of $J_{2}^{AA}$ and its minimum is along the $x$-direction. However, the non zero FB contribution behaves very differently. Below $\theta_1=\pi/4$ and above $\theta_2=3\pi/4$, $J_{1}^{BB}$ is ferromagnetic and antiferromagnetic for $\theta$ between these two specific angles. The largest antiferromagnetic coupling is almost ten times larger than the ferromagnetic one, as it was observed in the numerical calculations in the previous section. The maximum of $J_{1}^{BB}$ is reached at $\theta=\pi/2$, direction perpendicular to the (ABAB..) chains. 
In the case of a (B,C) pair, $J_{2}^{BC}$ and $J_{1}^{BC}$ have exactly the same $\theta$-dependence, but their sign differ, the first one is ferromagnetic while the other is antiferromagnetic and almost 4 times larger than the former. The details of the calculations can be found in the Appendix B.
Thus the B-C coupling is entirely  dominated bu the antiferromagnetic FB contribution.
\begin{table} [ht]
\caption{The $J^{BB}$ couplings in units of $J_0$} for $E_F=0$
 \centering
\begin{tabular}{c  c  c}  
\hline\hline
 \,\,\,\, \,\, \,\, \,$\theta$  \,\,\,\, \, \, \, \, &  Exact results \,  \,\,\, \, \, \, \, &  Analytical results \, \,\, \,\,\,  \\
 \hline 
 
 $0$   &   $J_1=+5\,10^{-5}a^{2}/R^{2}$     &      $J_1=0$    \\
          &    $J_2=-2.6 \,10^{-3}a^{3}/R^{3}$     &  $J_2= -2.5 \,10^{-3}a^{3}/R^{3}$      \\
\hline 
$\pi/4$  &   $J_1=0$     &      $J_1=0$    \\
          &     $J_2=-6.6 \,10^{-4}a^{3}/R^{3}$     &  $J_2= -6.2 \,10^{-4}a^{3}/R^{3}$      \\
\hline 
$\pi/2$   &   $J_1=+2.65\,10^{-2}a^{3}/R^{3}$     &      $J_1=+2.53 \,10^{-2}a^{3}/R^{3}$    \\
          &     $J_2=-5.6 \,10^{-3}a^{3}/R^{3}$     &   $J_2= -5.45 \,10^{-3}a^{3}/R^{3}$      \\

\end{tabular} 
\label{table:1}
\end{table}  

To estimate the validity of the linearization of the dispersive bands, we propose to compare the values found analytically with those obtained from the full numerical calculations. As it is illustrated in table \ref{table:1} for the B-B couplings, the agreement between the analytical calculations and the
full numerical calculations is very good, they differ by less than 5-6$\%$. Notice, for $\theta=0$, the discrepancy between the full calculations and the analytical result. Indeed, from the full calculation, $J_{1}^{BB}$ fall as off as $1/R^2$ while the analytical calculations predicts a vanishing coupling. The origin of this disagreement should be the linearization of the bands that leads only to the $1/R^3$ contribution. Thus one should include the quadratic term in the analytical calculations to obtain the $1/R^2$ decay.

To conclude this section devoted to the nature of the couplings at $E_F=0$, we observe that when we ignore the flat-band contribution contribution, the couplings between impurities belonging to the same sublattice are all ferromagnetic, including the B-C couplings and those between (A,B) and (A,C) are antiferromagnetic.
This would support a ferrimagnetic ordering of the spins, e.g. the magnetization of the localized spins on sublattice  A is the antiparallel to that of the two fold coordinated sites (B,C).
However, when we include the flat-band contribution the picture changes entirely. The couplings between (B,B) (resp. (C,C)) pairs are drastically affected by this additional contribution which is strongly antiferromagnetic for $\theta \in \left[ \pi/4,3\pi/4\right]$ (resp.  $\theta \in \left[ -\pi/4,\pi/4\right]$). Similarly, the couplings between (B,C) pairs, largely dominated 
by the FB term, switch to antiferromagnetic for any values of the angle $\theta$. Thus, when impurities are randomly distributed in the Lieb lattice, the $J_{1}^{\alpha\beta}$ contribution introduces frustration effects, the ferrimagnetic ground-state becomes unstable. Instead, we expect a spin-glass ground-state if the localized spins are randomly distributed on the two fold coordinated sites. Ferromagnetism is only possible if the magnetic impurities are distributed on the sublattice A only.

\section{Effects of the doping on the RKKY couplings}

In this last section, we propose to address the issue of the effects of tuning the Fermi energy. One can find the details of the analytical calculations in the Appendix D.
We first consider the case of the A-A couplings. 

For $x=k_FR \ll 1$ ($k_F=E_F/v_F$), where $v_F=a\,t$ is the Fermi velocity of the dispersive bands at the M point, we find,
\begin{eqnarray}
J^{AA,E_F}(\textbf{R})= J^{AA,0}(\textbf{R}) \Big(1+ \frac{32}{9\pi} (k_FR)^3 (1- \nonumber \\3\gamma-3\ln(\frac{k_FR}{2}))\Big),
\end{eqnarray}
where $\gamma=0.577$ is the Euler-Mascheroni constant. To obtain this expression, we have used the small $x$ expansion of $J_0(x)$ and $Y_0(x)$, the Bessel functions of first and second kind. Thus, the doping leads to a small $(k_FR)^3$ correction as it has also been found in the case of graphene.

On the other hand, in the opposite limit $k_FR \rightarrow \infty$, and using the asymptotic forms of $J_0(y)$ and $Y_0(y)$, we obtain,
\begin{eqnarray}
J^{AA,E_F}(\textbf{R})= \frac{2}{\pi}J^{AA,0}(\textbf{R}) \Big( 4k_FR\sin(2k_FR) \,\nonumber \\
 + \cos(2k_FR)       \Big).
\end{eqnarray}
Thus, for large distances and finite doping, the couplings are found to fall off as $1/R^2$ and oscillate with a period $\lambda_F= \frac{\pi}{k_F}$.
This characterizes the standard RKKY behavior in two dimensional systems. Notice, that these results are also similar to those found in the case of graphene \cite{sherafati2} since in the case of A-A couplings the flat-band has no impact.
\begin{figure}[t]\centerline
{\includegraphics[width=1.1\columnwidth,angle=0]{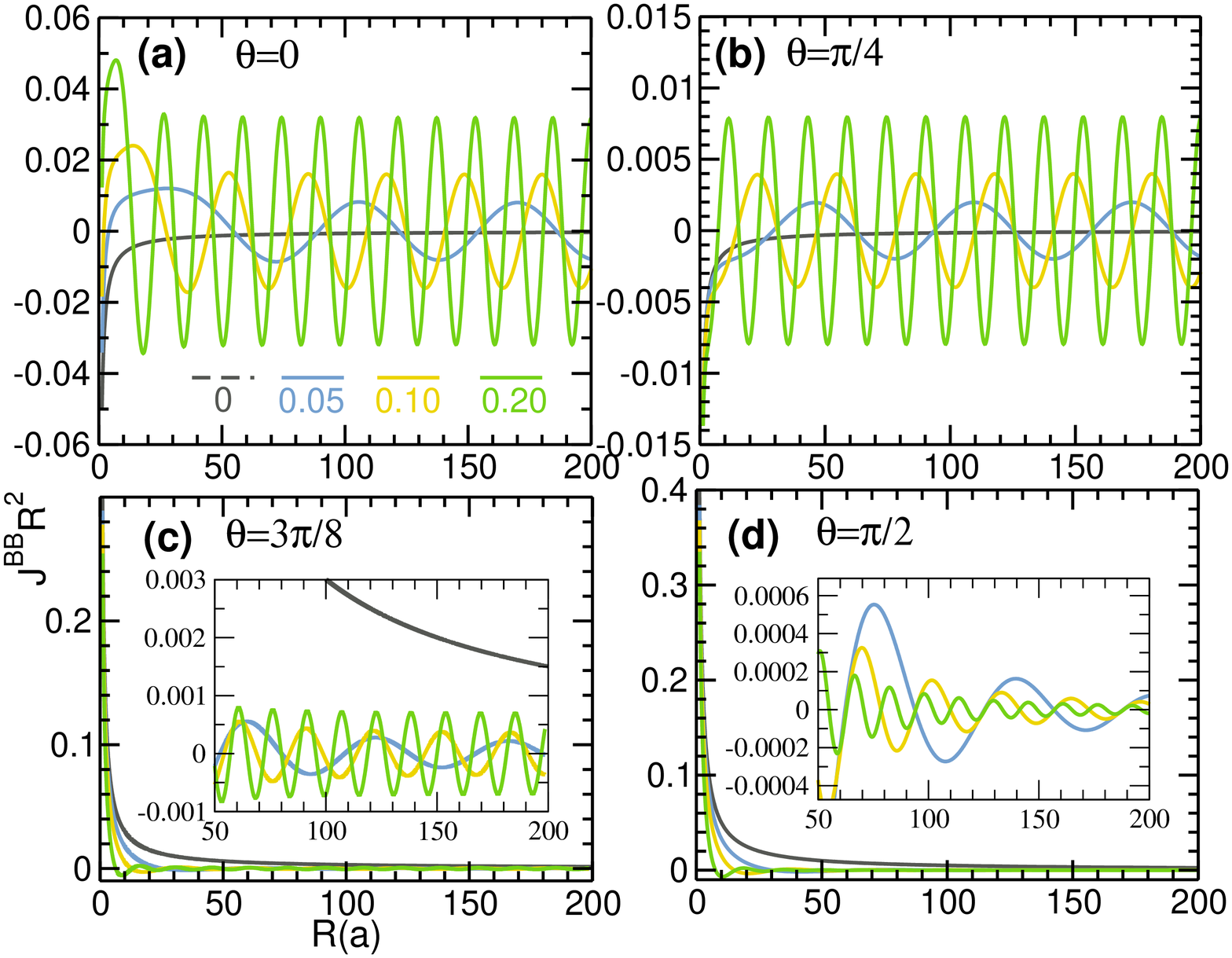}}
\caption{(Color online) 
$J_{BB}R^2$ as a function of R  for 4 different values of $\theta$ ($= 0$, $\pi/4$, $3\pi/8$ and $\pi/2$) where
$\theta$ is the angle between ${\bf R = R_{i}-R_{j}}$ (${\bf R_{i}}$ and ${\bf R_{j}}$ being the impurity positions) and the $x$-axis.
In each case ($\theta$ fixed), the Fermi energy $E_{F}$ is tuned: $E_{F}=0$ (dark grey), $0.05\,t$ (blue), $0.1\,t$ (yellow) and $0.2\,t$ (light green). 
The $J{BB}$ couplings are expressed in units of $J'_{0} = \frac{J^2}{8\pi t}$.} 
\label{fig6}
\end{figure} 

As it will be seen in what follows, in the case of the B-B (C-C and B-C) couplings the situation is very different. 
Let us start, with the $G^{BB}(\textbf{R},z)$ Green's function that can be put into the form,
\begin{eqnarray}
G^{BB}(\textbf{R},z)=-\frac{a^2}{2\pi} \frac{z}{v_F^2} \cos^{2}(\theta)
K_{0}(-i\frac{zR}{v_F}) + \,\nonumber \\
i\frac{a^{2}}{2\pi Rv_{F}}  \cos(2\theta)K_{1}(-i\frac{zR}{v_F}),
 \end{eqnarray}
where the Bessel functions of second kind $K_{0}$ and $K_{1}$ have been introduced. After re-expressing, them in terms of ordinary Bessel
 functions of first and second kind, we end up with (the details are also available in the Appendix D),
\begin{eqnarray}
J^{BB,E_F}(\textbf{R})= -\frac{J^2}{8\pi t} \frac{a^3}{R^3} \Big( \cos^4(\theta) F_{00}(k_FR) +   \nonumber  \\ 
\cos^2(2\theta) F_{11}(k_FR) - \cos(2\theta) \cos^2(\theta) F_{01}(k_FR) \Big).
 \label{jbbef}
 \end{eqnarray}
where we have now introduced a set of functions $F_{nl}$ ($n,l=0,1$) defined by,
\begin{eqnarray}
F_{nl}(k_FR) &= &\int_{k_F R}^{\infty}dy y^{2-n-l}  (J_n(y) Y_l(y)+ J_l(y) Y_n(y)) c_{nl},  \nonumber  \\ 
 \label{eqfi0}
 \end{eqnarray}
 where, $c_{nl}=1/2$ if $n=l$, otherwise $c_{nl}=1$.

In Fig.~\ref{fig6} are depicted the B-B couplings as a function of the distance between impurities for four different angles $\theta$ and four values of the Fermi energy $E_F$ as well. The couplings are directly obtained from the numerical calculation of the $F_{nl}$ integrals.
First, for a fixed angle $\theta$, we find a period of oscillations that decreases as $E_F$ increases and a $1/R^2$ fall off of the couplings. Second, we observe that the amplitude of the oscillations are drastically suppressed as the angle $\theta$ approaches $\pi/2$. For instance, for $E_F=0.1 t$, we find an amplitude of $0.016$, $0.004$ and $0.0004$
respectively for $\theta=0$, $\theta=\pi/4$ and $\theta=3\pi/4$, where the couplings is expressed in units of $J'_{0} = \frac{J^2}{8\pi t}$. To shed some light on these features observed in Fig.~\ref{fig6}, we propose to derive the analytical expression of the couplings in both limits $k_FR \ll 1$ (low doping/ short distance ) and $k_FR \gg 1$ (high doping/ long distance). Using the fact that $F_{00}(0)=1/16$, $F_{11}(0)=-1/2$ and $F_{01}(0)=-1/2$ (see Appendix D),one finds the analytical expression of the couplings for $k_FR \ll 1$,
\begin{eqnarray}
J^{BB,E_F}(\textbf{R})= J^{BB,0}(\textbf{R}) + \, \nonumber \\ 
\frac{J^2}{8\pi^2 t}\frac{a^3}{R^3} \cos(2\theta) k_FR +o((k_FR)^{3}).
\end{eqnarray}
Thus, for $R \le \lambda_F$, and in contrast to the A-A couplings, the correction to $J^{BB}(\textbf{R})$ due to the finite doping is linear in $k_F$ and $\theta$-dependent. Thus, the doping has a much stronger pronounced effect on the B-B (or C-C) couplings than on the A-A ones. This is again due to presence of the flat-band that directly contribute to the former couplings. One could check that the origin of this linear correction should be the $J_{1}^{BB}$ contribution, while $J_{2}^{BB}$ should give the standard $k_F^3$ correction .
Notice also that for the specific angles $\theta=\pi/4$ or $3\pi/4$ the linear correction vanishes.

Let us now consider the behavior of the B-B couplings in the opposite limit, $k_F R \gg 1$. Using both eqs.(\ref{jbbef}) and (\ref{eqfi}) and the large argument expansion of the Bessel functions $Y_i(y)$ and $J_i(y)$ ($i=0,1$), we find,
\begin{eqnarray}
J^{BB,E_F}(\textbf{R})= -\frac{J^2}{8\pi^2 t}\frac{a^3}{R^3} \Big( \frac{1}{2} \cos^4(\theta) x\sin(2x) + \, \nonumber \\ 
\frac{1}{8} \cos^2(\theta)(8-15\cos^2(\theta)) \cos(
2x)  \, \nonumber \\ 
+o(\frac{\sin(2x)}{x})+o(\frac{\cos(2x)}{x}) \Big),
\label{jbbinfty}
\end{eqnarray}
where $x=k_FR$.

Thus, for $\theta \le \frac{\pi}{2}$ the coupling fall off as $\frac{1}{R^2}$ and the amplitude of the oscillations appears to
scale as $\cos^4(\theta)$. This equation explains both the period of oscillations found for a fixed $\theta$ and the rapid drop of the amplitude as it has been observed in Fig.~\ref{fig6} when $\theta$ approaches $\pi/2$. For this particular angle, the first two terms on the right side of eq.(\ref{jbbinfty}) vanish, thus the couplings decay even more rapidly than in the undoped case. More precisely, along the $y$-axis direction, $J^{BB,E_F}$ falls off as $ \frac{1}{R^4}$. Notice again, that the doping effects on C-C couplings can be straightforwardly obtained by changing $\theta$ in $\frac{\pi}{2}-\theta$ in eq.(\ref{jbbinfty}).

 \section{Conclusion}

Motivated by the unconventional quantum electronic transport properties found in the Lieb lattice, we have addressed the effects of the flat band on the RKKY couplings. 
It has been shown that the presence of the flat band has drastic impacts in the couplings and lead to frustration effects when the magnetic impurities are lying in the two-fold coordinated sites. In addition, in contrast to the case of impurities on the four fold coordinated sites which is both ferromagnetic and isotropic, the coupling between impurities on the two-fold coordinated sites is strongly anisotropic and can be both ferromagnetic or anti-ferromagnetic. In all the cases, we have derived the analytical expressions of the couplings for both limits $k_F R \ll 1$ and $k_F R \gg 1$. We have found that the contribution to the couplings that originate from the matrix elements involving the flat band can dominate largely over the dispersive band contribution. The effects of tuning the Fermi energy has also revealed remarkable features.
It would be interesting, in the future to study RKKY couplings in other lattices, such as for instance, the Kagomé lattice where the flat band touches the
lower or upper dispersive band (depending on the sign of the nearest neighbor hopping) at the $\Gamma$ point. The introduction of disorder and the possibility of opening a gap between the FB and the dispersive bands will be investigated in the near future.

\begin{acknowledgments}
\end{acknowledgments}

\renewcommand{\theequation}{A.\arabic{equation}}
\setcounter{equation}{0}

\section*{Appendix A}
In this appendix, we provide the expression of the matrix elements needed for the calculation of the magnetic couplings. 
We first recall that 3 bands are associated to the Lieb lattice Hamiltionian (hoppings are restricted to nearest neighbor only): two are dispersive and third one is a flat-band. Their respective energy are,
\begin{eqnarray}
E^\pm({\bf k})=\pm 2t f_0({\bf k}), \\ \,
E^0({\bf k})=0.
 \end{eqnarray}
where $f_0({\bf k})=\sqrt{\cos^2(k_xa/2)+\cos^2(k_ya/2)}$. The associated eigenvectors are,
\begin{eqnarray}
\vert \Psi ^\pm_{\bf k}\rangle &=& \frac{\pm F_x({\bf k}) \vert B{\bf k} \rangle \pm F_y({\bf k}) \vert C{\bf k} \rangle  +  
\vert A{\bf k} \rangle }{\sqrt{2}}, \\ \,
\vert \Psi ^0_{\bf k}\rangle &=& -F_y({\bf k})) \vert B{\bf k} \rangle + F_x({\bf k})) \vert C{\bf k} \rangle, 
 \end{eqnarray}
 where $F_\lambda({\bf k})) = cos(k_\lambda a/2)/f_0({\bf k})$ ($\lambda=x,y$) and $\vert \alpha{\bf k} \rangle =\frac{1}{\sqrt{N}} \sum_{\bf i}  e^{\bf k.R_i}   \vert \alpha {\bf i} \rangle, $ ($\alpha= A,B$ and $C$), $N$ is the number of unit cells.

Find below the list of the matrix elements defined by,
$C_{\alpha\beta}^{\lambda\lambda'} ({\bf k},{\bf k'},
{\bf R})=\langle {\bf i} \alpha \vert  \Psi^\lambda_{\bf k}\rangle \langle  \Psi ^\lambda_{\bf k} \vert \beta {\bf j}  \rangle
\langle {\bf j} \beta \vert  \Psi^ {\lambda' }_{\bf k'}\rangle \langle  \Psi ^{\lambda'}_{\bf k'} \vert \alpha {\bf i}  \rangle$, where $\alpha$ and $\beta$ are the  sublattice index, and
$\lambda$ and $\lambda'$ the band index:
  \begin{eqnarray}
  C_{AA}^{+-} ({\bf k},{\bf k'},{\bf R})&=&\frac{1}{4N^2} e^{i\textbf{q} \cdot \textbf{R}}, \,\nonumber \\   
  C_{AA}^{0+} ({\bf k},{\bf k'},{\bf R})&=&0, 
  \end{eqnarray}
  \begin{eqnarray}
  C_{BB}^{+-} ({\bf k},{\bf k'},{\bf R})&=&\frac{1}{4N^2} F^2_x({\bf k}) F^2_x({\bf k'})  e^{i\textbf{q}\cdot\textbf{R}}, \,\nonumber  \\
  C_{BB}^{0+} ({\bf k},{\bf k'},{\bf R})&=&\frac{1}{2N^2} F^2_y({\bf k}) F^2_x({\bf k'})  e^{i\textbf{q}\cdot\textbf{R}}, 
  \end{eqnarray}  
\begin{eqnarray}   
C_{AB}^{+-} ({\bf k},{\bf k'},{\bf R})&=&-\frac{1}{4N^2} F_x({\bf k}) F_x({\bf k'})  e^{i\textbf{q}\cdot\textbf{R}},  \,\nonumber \\
C_{AB}^{0+} ({\bf k},{\bf k'},{\bf R})&=&0,
\end{eqnarray}  
\begin{eqnarray}   
C_{BC}^{+-} ({\bf k},{\bf k'},{\bf R})&=&\frac{1}{4N^2} F_x({\bf k}) F_y({\bf k})  F_x({\bf k'})F_y({\bf k'})  e^{i\textbf{q}\cdot\textbf{R}}, \, \nonumber  \\
 C_{BC}^{0+} ({\bf k},{\bf k'},{\bf R})&=&
-\frac{1}{2N^2} F_x({\bf k}) F_y({\bf k})  \times \, \nonumber  \\
F_x({\bf k'})F_y({\bf k'})e^{i\textbf{q}\cdot\textbf{R}},
\end{eqnarray}  
where, $\textbf{R}=\textbf{R}_i-\textbf{R}_j$ and $\textbf{q}=\textbf{k}-\textbf{k'}$.
The missing $C_{AC}^{\lambda\lambda'}$ and $C_{CC}^{\lambda\lambda'} $can be straightforwardly calculated.

\renewcommand{\theequation}{B.\arabic{equation}}
\setcounter{equation}{0}

\section*{Appendix B}

In this appendix, we derive the analytical expressions of the two different contributions to the couplings between pairs of magnetic impurities: the inter-dispersive band term and the FB contribution.

The exchange between two impurities located respectively on a site of type $\alpha$ and the other on a $\beta$-type reads, 
\begin{eqnarray}
J_{\bf ij}^{\alpha\beta} =-2J^2 \huge( \sum_{\bf k k'} \frac {1}{\epsilon^0({\bf k} )} C_{\alpha\beta}^{0+} ({\bf k},{\bf k'},
{\bf R})\, \nonumber \\
+ \sum_{\bf k k'}  \frac {1}{\epsilon^0({\bf k}) +\epsilon^0({\bf k'} )}  C_{\alpha\beta}^{+-} ({\bf k},{\bf k'},
{\bf R})
\huge).
\label{coupling3}
\end{eqnarray}
The first contribution is associated to the matrix elements that involve the flat-band and the conduction band. The second term originates from matrix elements between the valence band and the conduction band. We define $J_{1}^{\alpha \beta} (\bf R)$ (resp.  $J_{2}^{\alpha \beta} (\bf R)$ ) as the first (resp. second) term in eq.(\ref{coupling3}).
To allow the analytical calculations of the couplings, we first linearize the dispersions in the vicinity of the M=$(\pi,\pi)$ point in the Brillouin zone.
Let us first consider the case of the (B,B) pair of magnetic impurities,
\begin{eqnarray}
J_{1}^{BB} ({\bf R})  =-\frac{J^2 }{t}F_{a}({\bf R}) F_{b}({\bf R}),
\end{eqnarray}
where,
\begin{eqnarray}
F_{a}({\bf R})=e^{i{\bf \pi}\cdot\textbf{R}} \frac{a^2}{4\pi^2} \int_0^{q_c} qdq \int_{-\pi}^{+\pi}  d\theta_q \sin^2(\theta_q )e^{iqR\cos(\theta_{q,R})}, \, \nonumber \\  
F_{b}({\bf R})=e^{-i{\bf \pi}\cdot\textbf{R}} \frac{a^2}{4\pi^2} \int_0^{q_c}dq \int_{-\pi}^{+\pi}d\theta_q \cos^2(\theta_q )e^{iqR\cos(\theta_{q,R})} , \, \nonumber \\ 
\end{eqnarray}
$\theta_{q,R}$ is the angle between $\textbf{q}$ and \textbf{R}. $q_c$ is the cut-off introduced after the linearization of the dispersive bands, in what follows we set 
$q_c =\infty$, which is appropriate to describe the long distance behavior of the couplings. 
In the next step, we introduce the Bessel  functions of first kind, $J_0(x)$ and $J_1(x)$, and after performing a change of variables, we can re-express $F_a({\bf R})$ and 
$F_b({\bf R})$ as follows,
\begin{eqnarray}
F_{a}({\bf R})=\frac{a^2}{2\pi R^2} e^{i{\bf \pi}\cdot\textbf{R}} \Big( \cos^2(\theta) I_1 + \cos(2\theta) I_2 \Big), \, \nonumber \\
F_{b}({\bf R})=\frac{a}{2\pi R} e^{-i{\bf \pi}\cdot\textbf{R}} \Big(\sin^2(\theta) I_3  \, \nonumber \\
-\cos(2\theta) (-I_3 +I_4)\Big),   
\end{eqnarray}
where $I_1=\int_{0}^{+\infty} J_0(u) u du $, $I_2=\int_{0}^{+\infty} J_1(u) du $,  $I_3=\int_{0}^{+\infty} J_0(u) du $ and $I_4= \int_{0}^{+\infty}(J_1(u)/u) du $.
It can be shown that $I_1=0$ and $I_2=I_3=I_4=1$ \cite{book-integrals}. Notice that we have also used the fact that,
 \begin{eqnarray}
 \frac{d^2J_0}{du^2} =-J_0(u)+\frac{J_1(u)}{u}.
 \end{eqnarray} 
Thus, we finally find for the FB contribution,
\begin{eqnarray}
J_{1}^{BB} ({\bf R}) = -\frac{1}{4\pi^2} \frac{a^3}{R^3} \sin^2(\theta)  \cos(2\theta) \frac{J^2}{t}.
\end{eqnarray}

We now proceed further with the case of $J_{2}^{BB}$. After linearization of the dispersion of the valence and conduction bands, one obtains,
\begin{eqnarray}
J_{2}^{BB} ({\bf R}) = -\frac{a^3 J^2}{32\pi^4 t} \int \frac{qq'}{q+q'}dqdq' d\theta_q d\theta_{q' }  \,  \nonumber  \\ 
\cos^2(\theta_q) \cos^2(\theta_{q'})e^{iqR\cos(\theta_{q,R})} e^{iq'R\cos(\theta_{{q'},R})} .
 \end{eqnarray}
 Here, the single $\int$ sign means that q and q' are integrated from 0 to $\infty$ and $\theta_q$ and $\theta_{q'}$ from $-\pi$ to $\pi$.
 As before, the introduction of the Bessel functions leads to,
 \begin{eqnarray}
J_{2}^{BB} ({\bf R}) =-\frac{a^3J^2}{8\pi^2 R^3 t} \int dudv\frac{uv}{u+v} \,  \nonumber  \\ 
(A_\theta J_0(u)+B_\theta \frac{J_1(u)}{u})(A_\theta J_0(v)+B_\theta\frac{J_1(v)}{v}) \,  \nonumber  \\ 
=-\frac{a^3J^2}{8\pi^2 R^3 t} \sum_{n,p=0}^{n,p=1} I_{n,p} (A_\theta)^{2-n-p}(B_\theta)^{n+p},
 \end{eqnarray}
 where $A_\theta=\cos^2(\theta)$ and $B_\theta=-\cos(2\theta)$ and,
 \begin{eqnarray}
I_{n,p}=\int_{0}^{\infty}\int_{0}^{\infty} dudv\frac{u^{1-n}v^{1-p}}{u+v} J_{n}(u)J_{p}(v).
\label{inp}
 \end{eqnarray}
The calculation of these integrals can be found in the appendix C. The dispersive band contribution finally reads,
 \begin{eqnarray}
J_{2}^{BB} ({\bf R}) =-\frac{a^3J^2}{8\pi^2 R^3 t} \big( (\frac{\pi}{16} +4C)\cos^4(\theta)+\,  \nonumber  \\ 
2C(-3\cos^2(\theta)+1) \big),
 \end{eqnarray}
 where $C=1-\frac{\pi}{4}$. 
 
 Following this procedure, we can derive the expression of the other couplings as well,
 \begin{eqnarray}
J_{2}^{AA} ({\bf R}) =-\frac{1}{128\pi}\Big(\frac{a}{R}\Big)^3 \frac{J^2}{t},
\label{j2aa}
 \end{eqnarray} 
 \begin{eqnarray}
J_{2}^{AB} ({\bf R}) =\frac{3}{128\pi}\Big(\frac{a}{R}\Big)^3\cos^2(\theta)\frac{J^2}{t}, 
\end{eqnarray} 
\begin{eqnarray}
J_{1}^{BC} ({\bf R}) =\frac{1}{8\pi^2}\Big(\frac{a}{R}\Big)^3\sin^2(2\theta)\frac{J^2}{t}, \nonumber  \, \\
J_{2}^{BC} ({\bf R}) =-\frac{1}{32\pi^2}(4C+\frac{\pi}{16})
\Big(\frac{a}{R}\Big)^3\sin^2(2\theta)\frac{J^2}{t}, \nonumber  \, \\
 \end{eqnarray} 
In addition, $J_{1}^{AA} ({\bf R}) =0 $ and J$_{1}^{AB} ({\bf R}) =0$, since there is no flat band contribution if one of the magnetic impurity site is of type A.

\renewcommand{\theequation}{C.\arabic{equation}}
\setcounter{equation}{0}

\section*{Appendix C}

 This appendix is devoted to the analytical expressions of the following  set of integrals,
 \begin{eqnarray}
J^{nl}_{\mu\nu}=\int_{0}^{\infty}\int_{0}^{\infty} dudv\frac{u^{n}v^{l}}{u+v} J_{\mu}(u)J_{\nu}(v).
 \end{eqnarray} 
As it has been done in the case of graphene \cite{sherafati3}, we first define the following function with $s\ge 0$,
 \begin{eqnarray}
F(s)=\int_{0}^{\infty}\int_{0}^{\infty} dudv e^{-s(u+v)} \frac{u^{n} v^{l} }{u+v}J_{\mu}(u) J_{\nu}(v). \nonumber \,\\
 \end{eqnarray} 
 Its derivative can be written,
 \begin{eqnarray}
\frac{dF(s)}{ds}=-\mathcal{L}(x^{n}J_{\mu}(x)).\mathcal{L}(x^{l}J_{\nu}(x)),\,
\label{eqa} 
 \end{eqnarray} 
 where the Laplace transform is defined by,
 \begin{eqnarray}
\mathcal{L}(J_{\mu}(x))= \int_{0}^{\infty} dx e^{-sx} J_{\mu}(x).
 \end{eqnarray} 
According to Ref. \cite{book-integrals} the Laplace transform can be put into the form,
 \begin{eqnarray}
\mathcal{L}(J_{\mu}(x))= \frac{(\sqrt{s^2-1}-s)^{\mu}}{\sqrt{s^2+1}}.
\label{eqb} 
 \end{eqnarray} 
Thus,
 \begin{eqnarray}
\mathcal{L}(x^{n}J_{\mu}(x))= (-1)^{n} \frac{d^n}{ds^{n}} \Big(  \frac{(\sqrt{s^2-1}-s)^{\mu}}{\sqrt{s^2+1}}  \Big).
 \end{eqnarray} 
 As a first case, we can now calculate $I_{1,0} $. In eq. (\ref{eqa}), we set $\mu=0$, $\nu=1$, $n=1$ and $l=0$. 
 Using  (\ref{eqb}) we find,
 \begin{eqnarray}
 \frac{dF(s)}{ds}=-\frac{s}{(s^2+1)^{3/2}} + \frac{s^2}{(s^2+1)^{2}} .
  \end{eqnarray} 
Since $\lim_{s\rightarrow \infty }F(s) =0$, the integration of this equation leads to,
  \begin{eqnarray}
 F(s)=\frac{1}{(s^2+1)^{1/2}} -\frac{1}{2}\frac{s}{(s^2+1)} +\,  \nonumber  \\
  \frac{1}{2} \tan^{-1}(s) -\frac{\pi}{4}.
  \end{eqnarray} 
We finally obtain,
 \begin{eqnarray}
J^{10}_{01}= F(0)=I_{01}= I_{10}= 1-\frac{\pi}{4},
 \end{eqnarray} 
 where the integral $I_{np}$ is defined in Eq. (\ref{inp}). Following this procedure, we also find,
\begin{eqnarray}
I_{00}=\frac{\pi}{16}, \, \\
I_{11}=2 \,I_{01}.
 \end{eqnarray}

\renewcommand{\theequation}{D.\arabic{equation}}
\setcounter{equation}{0}
\section*{Appendix D}
 
 In this appendix, we address the effects of electron/hole doping in the couplings.
 We first consider the case of (A,A) couplings. We start with the Green's function $G^{AA}(\textbf{R},z)$ that can be written,
 \begin{eqnarray}
G^{AA}(\textbf{R},z)=\frac{za^2}{2\pi}\int_{0}^{\infty}qdq \frac{1}{z^2-v^{2}_{F}q^{2}}J_{0}(qR),
\end{eqnarray}
where $z=E+i\eta$, $v_F=a\,t$ is the Fermi velocity of the dispersive bands at the M point (we set $\hbar=1)$.
Using the fact that, for both $\alpha> 0$ and $\Re(k) > 0$ \cite{book-integrals},
 \begin{eqnarray}
 \int_{0}^{\infty}dx \frac{xJ_0(\alpha x)}{x^2+k^2}=K_0(\alpha k).
 \label{eqK0}
 \end{eqnarray}
where $K_{0}$ is the modified Bessel function of second kind. Then, we can write,
\begin{eqnarray}
G^{AA}(\textbf{R},z)=-\frac{a^2}{2\pi} \frac{z}{v_F^2} K_{0}(-i\frac{zR}{v_F}).
\end{eqnarray}
We now re-express $K_{0}$ in terms of $J_{0}$ and $Y_{0}$, the ordinary Bessel functions of first and second kind,
\begin{eqnarray}
K_{0}(-iz)=\frac{\pi}{2} (iJ_0(z)-Y_0(z)).
\label{joyo}
\end{eqnarray}
Thus, one finds,
\begin{eqnarray}
\Im(G^{AA}(\textbf{R},z)^2)
=-\frac{1}{8t^2} \frac{a^2}{R^2} y^2J_0(y)Y_0(y),
 \label{gaa2}
 \end{eqnarray}
where we have introduced the reduced variable $y=\frac{RE}{v_F}$. 
Thus, the couplings between (A,A) pairs for a fixed Fermi energy can be writtren,
\begin{eqnarray}
J^{AA,E_F}(\textbf{R})= J^{AA,0}(\textbf{R}) +  \nonumber \\
\frac{J^2}{8\pi t} \frac{a^3}{R^3} \int_{0}^{k_FR} dy y^2J_0(y)Y_0(y),
\label{jaa-ef}
\end{eqnarray}
where $J^{AA,0}(\textbf{R})$ is the coupling calculated in the appendix B at $E_F=0$. 
To obtain $J^{AA,E_F}$ for $k_F R \ll 1$, we can use the small $y$ expansion of $J_0(y)$ and $Y_0(y)$, this leads to,
 \begin{eqnarray}
 y^2J_0(y)Y_0(y)=\frac{2}{\pi} \Big( \ln(\frac{y}{2}) y^{2} + \gamma y^2 \Big) +o(y^4),
\end{eqnarray}
where $\gamma=0.577$ is the Euler-Mascheroni constant. After few straightforward steps, we obtain,
 \begin{eqnarray}
J^{AA,E_F}(\textbf{R})= J^{AA,0}(\textbf{R}) \Big(1+ \frac{32}{9\pi} (k_FR)^3 (1- \nonumber \\3\gamma-3\ln(\frac{k_FR}{2}))\Big).
\end{eqnarray}

Now, to derive the couplings in the opposite limit, $k_F R \gg 1$. We first re-express eq. (\ref{jaa-ef}) in the following form,
 $J^{AA,E_F}(\textbf{R})=\frac{J^2}{8\pi t} \frac{a^3}{R^3} \int_{k_FR}^{\infty} dy y^2J_0(y)Y_0(y)$ and use the asymptotic forms of $J_0(y)$ and $Y_0(y)$,
 \begin{eqnarray}
J_0(x)=\sqrt{\frac{2}{\pi}}\Big(\frac{\cos(x-\frac{\pi}{4})}{x^{1/2}} + \frac{1}{8} \frac{\sin(x-\frac{\pi}{4})}{x^{3/2}} +o(1/x^{5/2} \Big),  \, \nonumber \\
Y_0(x)=\sqrt{\frac{2}{\pi}}\Big(\frac{\sin(x-\frac{\pi}{4})}{x^{1/2}} - \frac{1}{8} \frac{\cos(x-\frac{\pi}{4})}{x^{3/2}} +o(1/x^{5/2} \Big). \, \nonumber \\
\label{j0-y0}
 \end{eqnarray} 
 We finally get, 
\begin{eqnarray}
J^{AA,E_F}(\textbf{R})= \frac{2}{\pi}J^{AA,0}(\textbf{R}) \Big( 4k_FR\sin(2k_FR) \,\nonumber \\
+ \cos(2k_FR)   \Big).
\end{eqnarray}
Notice that to calculate properly the large $k_F R$ expression of $J^{AA,E_F}$, it is essential to include in the expansion of $J_0(x)$ and $Y_0(x)$ the $1/x^{3/2}$ terms.  

We now proceed further and consider the case of $J^{BB,E_F}$. Starting with the definition of $G^{BB}(\textbf{R},z)$, and after few elementary steps, one can write,
 \begin{eqnarray}
G^{BB}(\textbf{R},z)= I_1 + I_2,
\end{eqnarray} 
where, after performing the integration over $\theta_q$, one finds,
 \begin{eqnarray}
I_1=\frac{z a^2}{2\pi}\int_{0}^{\infty} dq \frac{q}{z^{2}-(v_Fq)^{2}} f_1(q,\theta),
 \nonumber \\
 I_2= \frac{a^2}{2\pi z}\int_{0}^{\infty} qdqf_2(q,\theta),
\end{eqnarray}  
where, $f_1(q,\theta)=cos^2(\theta) J_0(qR) - \cos(2\theta) \frac{J_1(qR)}{qR} $ and 
 $f_2(q,\theta)=\sin^2(\theta) J_0(qR) + \cos(2\theta) \frac{J_1(qR)}{qR}$.
 
From eq.(\ref{eqK0}) and its derivative with respect to $\alpha$, we end up with the following expression for $G^{BB}(\textbf{R},z)$,
 \begin{eqnarray}
G^{BB}(\textbf{R},z)=-\frac{a^2}{2\pi} \frac{z}{v_F^2} \cos^{2}(\theta)
K_{0}(-i\frac{zR}{v_F}) + \,\nonumber \\
i\frac{a^{2}}{2\pi Rv_{F}}  \cos(2\theta)K_{1}(-i\frac{zR}{v_F}).\,\nonumber \\
 \end{eqnarray}
 Using eq.(\ref{joyo}) and the fact that $K_{1}(-iz)=-\frac{\pi}{2} (iJ_1(z)+Y_1(z))$, we obtain for the calculation of the B-B couplings,
 \begin{eqnarray}
\Im(G^{BB}(\textbf{R},z)^2)
=-\frac{\pi^2}{2} \Big(  \alpha^2 J_0(z_1)Y_0(z_1) E^2  + \beta^2 J_1(z_1)Y_1(z_1) \nonumber \\
-\alpha \beta ( J_0(z_1)Y_1(z_1) +  J_1(z_1)Y_0(z_1) E \Big), \nonumber \\
\end{eqnarray} 
 where the new variable $z_1=\frac{zR}{v_F}$, and the $\theta$ dependent functions are respectively $\alpha=-\frac{1}{2\pi v^{2}_F}\cos^2(\theta)$ and $\beta=\frac{1}{2\pi v_F}\cos(2\theta)$. 
 We finally obtain,  
  \begin{eqnarray}
J^{BB,E_F}(\textbf{R})= -\frac{J^2}{8\pi t} \frac{a^3}{R^3} \Big( \cos^4(\theta) F_{00}(k_FR) +   \nonumber  \\ 
\cos^2(2\theta) F_{11}(k_FR) - \cos(2\theta) \cos^2(\theta) F_{01}(k_FR) \Big). \nonumber  \\ 
 \label{jbbef}
 \end{eqnarray}
We have introduced $F_{nl}$ functions ($n,l=0,1$) that are defined as follows,
\begin{eqnarray}
F_{nl}(k_FR) &= &\int_{k_F R}^{\infty}dy y^{2-n-l}  (J_n(y) Y_l(y)+ J_l(y) Y_n(y)) c_{nl},  \nonumber  \\ 
 \label{eqfi}
 \end{eqnarray}
 where, $c_{nl}=1/2$ if $n=l$, otherwise $c_{nl}=1$.
 
For the small doping behavior of $J^{BB,E_F}$, one needs the values of $F_{nl}(0)$, that are respectively, $F_{00}(0)=1/16$, $F_{11}(0)=-1/2$ and $F_{01}(0)=-1/2$. 
The details of the calculations are available in the Appendix E. With these values of $F_{nl}(0)$ and the small argument expansion of $J_n(y)$ and $Y_l(y)$, we find,
\begin{eqnarray}
J^{BB,E_F}(\textbf{R})= J^{BB,0}(\textbf{R}) + \, \nonumber \\ 
\frac{J^2}{8\pi^2 t}\frac{a^3}{R^3} \cos(2\theta) k_FR +o((k_FR)^{3}).
\end{eqnarray}
   
We now calculate the B-B couplings in the opposite limit, e.g. $k_F R \gg 1$. For large values of the argument $x$,
 \begin{eqnarray}
J_1(x)=\sqrt{\frac{2}{\pi}}\Big(\frac{\cos(x-\frac{3\pi}{4})}{x^{1/2}} -\frac{3}{8} \frac{\sin(x-\frac{3\pi}{4})}{x^{3/2}} +o(1/x^{5/2} \Big),  \,  \nonumber \\
Y_1(x)=\sqrt{\frac{2}{\pi}}\Big(\frac{\sin(x-\frac{3\pi}{4})}{x^{1/2}} +\frac{3}{8} \frac{\cos(x-\frac{3\pi}{4})}{x^{3/2}} +o(1/x^{5/2} \Big). \, \nonumber \\
 \end{eqnarray} 
 The large argument expansion of $J_0(x)$ and $Y_0(x)$ are already given in Eq.(\ref{j0-y0}). 
 Finally, combining both Eq.(\ref{jbbef}) and Eq.(\ref{eqfi}) leads to,
 \begin{eqnarray}
J^{BB,E_F}(\textbf{R})= -\frac{J^2}{8\pi^2 t}\frac{a^3}{R^3} \Big( \frac{1}{2} \cos^4(\theta) x\sin(2x) + \, \nonumber \\ 
\frac{1}{8} \cos^2(\theta)(8-15\cos^2(\theta)) \cos(
2x)  \, \nonumber \\ 
+o(\frac{\sin(2x)}{x})+o(\frac{\cos(2x)}{x}) \Big). \, \nonumber \\
\end{eqnarray}

\renewcommand{\theequation}{E.\arabic{equation}}
\setcounter{equation}{0}
\section*{Appendix E}
 In this appendix we calculate the values of $F_{nl}(0)$ as defined in Eq.( \ref{eqfi}). From the expression of $J^{AA,0}$ as it is given in Eq.( \ref{j2aa}) and the other one that could be directly derived from Eq.( \ref{gaa2}), one can unambiguously conclude that $F_{00}(0)=1/16$.
 
Using the fact that \cite {book-integrals},
 \begin{eqnarray}
\int_{0}^{\infty}dy J_n(ay) Y_n(ay)=-\frac{1}{2a},
\label{e1}
 \end{eqnarray}
one immediately gets,
 \begin{eqnarray}
\int_{0}^{\infty}dy y (J_0(ay) Y_1(ay)+J_1(ay) Y_0(ay)) =-\frac{1}{2a^2} \, \nonumber \\
\end{eqnarray}
 where, this equation is simply obtained by the derivation of Eq. (\ref{e1}) with respect to a. 
 Thus, we conclude that $F_{11}(0)=-1/2$ and $F_{01}(0)=-1/2$, both needed for the calculation of the couplings.

\newpage


\begin{thebibliography}{99}





\bibitem{julku} A. Julku, S. Peotta, T. I. Vanhala, D.-H. Kim, and P. Torma, Phys. Rev. Lett. \textbf{117}, 045303 (2016).
\bibitem{peri} V.Peri, Zhi-Da Song, B. A. Bernevig, and S. D. Huber, Phys. Rev. Lett. \textbf{126}, 027002 (2021).
\bibitem{peotta}S. Peotta and P. Torma, Nature Comm. \textbf{6}, 8944 (2015.)

\bibitem{lizardo} L. H. C. M. Nunes and C. M. Smith, Phys. Rev. B \textbf{101}, 224514 (2020).
\bibitem{miyahara}S. Miyahara, S. Kusuta, and N. Furukawa, Physica C: Superconductivity \textbf{460,} 1145 (2007).
\bibitem{aoki}H. Aoki , Journal of Superconductivity and Novel Magnetism,\textbf{33,} 2341(2020).



\bibitem{cao} Y. Cao, V. Fatemi, A. Demir, S. Fang, S. L. Tomarken, J.Y. Luo, J.D. Sanchez-Yamagishi, K. Watanabe, T. Taniguchi, E. Kaxiras, R.C. Ashoori and P. Jarillo-Herrero, Nature \textbf{556}, 43 (2018).

\bibitem{chen1} 
G. Chen, A. L. Sharpe, P. Gallagher, I.T. Rosen, E. J. Fox, L. Jiang, B. Lyu, H. Li, K. Watanabe, T. Taniguchi, J. Jung, Z. Shi, D. Goldhaber-Gordon, 
Y. Zhang and F. Wang ,Nature \textbf{572,} 215, (2019).

\bibitem{yankowitz} 
M. Yankowitz, S. Chen, H. Polshyn, Y. Zhang, K. Watanabe, T. Taniguchi, D. Graf, A. F. Young,  and C. R. Dean, Science \textbf{363}, 1059 (2019).

\bibitem{park} J. Min Park, Y. Cao, K. Watanabe, T. Taniguchi and P. Jarillo-Herrero, Nature \textbf{ 590,} 249 (2021).




\bibitem{lieb} E.H. Lieb Phys. Rev. Lett., {\bf 62}, 1201 (1989).
\bibitem{tasaki}  H. Tasaki, Phys. Rev. Lett. \textbf{ 69} 1608 (1992); Prog. Theor. Phys. \textbf{ 99}(4):\textbf{489} (1998).
\bibitem{mielke}  A. Mielke, Phys. Rev. Lett. \textbf{82}, 4312 (1999).
\bibitem{noda} K. Noda, A. Koga, N. Kawakami, and T. Pruschke, Phys. Rev. A \textbf{80}, 063622 (2009)`
\bibitem{jiang} W.Jiang, H. Huang and F. Liu 
Nature Comm. \textbf{10}, 2207 (2019).

\bibitem{venkatesan}M. Venkatesan, C. B. Fitzgerald and J. M.D. Coey, Nature (London) \textbf{430,} 630 (2004).
\bibitem{young1}D. P. Young et al., Nature (London) \textbf{397}, 412 (1999)
\bibitem{makarova}T. L. Makarova et al., Nature (London) \textbf{413,} 716 (2001).
\bibitem{bouzerard0}G. Bouzerar and T. Ziman, Phys. Rev. Lett. \textbf{96}, 207602 (2006).




\bibitem{kang}Topological flat bands in frustrated kagome lattice CoSn
M. Kang, S. Fang, L. Ye, H. Chun Po, J. Denlinger, C. Jozwiak, A. Bostwick, E. Rotenberg, E. Kaxiras, J. G. Checkelsky and R. Comin 
Nature Comm. \textbf{ 11,} 4004 (2020).

\bibitem{gao}Z. Gao and Z. Lan, Phys. Rev. B \textbf{102,} 245133 (2020).


\bibitem{liu} X. Liu, C.-Li Chiu, J. Y. Lee, G. Farahi, K. Watanabe, T. Taniguchi, A.Vishwanath and A. Yazdani, Nature Comm. \textbf{12}, 2732 (2021).
\bibitem{tang} E. Tang, J.-W. Mei, and X.-G. Wen, Phys. Rev. Lett. \textbf{106}, 236802 (2011).
\bibitem{sun} K. Sun, Z. Gu, H. Katsura, and S. Das Sarma, Phys. Rev. Lett. \textbf{106}, 236803 (2011).



\bibitem{pal} B Pal, Phys. Rev. B \textbf{98,} 245116 (2018). 

\bibitem{neupert}T. Neupert, L. Santos, C. Chamon, and C. Mudry, Phys. Rev. Lett. \textbf{106}, 236804 (2011).

\bibitem{wu1}C. Wu, D. Bergman, L. Balents, and S. Das Sarma, Phys. Rev. Lett. \textbf{99}, 070401 (2007).

\bibitem{chen} Y. Chen, S. Xu, Y. Xie, C. Zhong, C.Wu, and S. B. Zhang, Phys. Rev. B \textbf{98,} 035135 (2018).

\bibitem{bouzerarSM12} G. Bouzerar and D. Mayou, Phys. Rev. B \textbf{103}, 075415 (2021).


\bibitem{hase}I. Hase, T. Yanagisawa and K. Kawashima,  Nanoscale Research Letters \textbf{13}, 63 (2018).
\bibitem{xu}
C. Xu, G. Wang, Z. H. Hang, J. Luo, C. T. Chan and Y. Lai , Scientific Reports \textbf{5}, 18181 (2015) 
\bibitem{mizoguchi} T. Mizoguchi and M.Udagawa Phys. Rev. B \textbf{99}, 235118 (2019). 
\bibitem{vergel}
N. A. Franchina Vergel, L.C. Post, D. Sciacca, M. Berthe, F. Vaurette, Y. Lambert, D. Yarekha, D. Troadec, C. Coinon, G. Fleury, G. Patriarche, T. Xu, L. Desplanque, X. Wallart, D. Vanmaekelbergh, C. Delerue, and B. Grandidier, Nano Lett.  \textbf{21}, 1, 680–685 (2021).
\bibitem{kerelsky} A. Kerelsky, C. Rubio-Verdú, L. Xian, D. M. Kennes,  D. Halbertal, N. Finney, L. Song, S. Turkel, L. Wang, K. Watanabe, T. Taniguchi, J. Hone, C. Dean, D.N. Basov,  A. Rubio, and A.N. Pasupathy, PNAS \textbf{118 } No. 4 e2017366118 (2021).




\bibitem{ruderman}M. A. Ruderman and C. Kittel, Phys. Rev. \textbf{96}, 99 (1954).
\bibitem{kasuya} T. Kasuya, Prog. Theor. Phys. \textbf{16}, 45 (1956).
\bibitem{yosida} Phys. Rev. \textbf{106}, 893 (1957).



 \bibitem{rkky-coupling} M. I. Katsnelson and A. I. Lichtenstein, Phys. Rev. B \textbf{61}, 8906 (2000).
 
 
\bibitem{sherafati1} M. Sherafati and S. Satpathy, Phys. Rev. B \textbf{83}, 165425 (2011).
\bibitem{sherafati2}M. Sherafati and S. Satpathy, Phys. Rev. B \textbf{84}, 125416 (2011).
\bibitem{saremi} S. Saremi, Phys. Rev. B \textbf{76}, 184430 (2007).
\bibitem{kogan} E. Kogan, Phys. Rev. B \textbf{84}, 119909 (2011).


\bibitem{book-integrals} S. Gradshteyn and I. M. Ryzhik, Tables of Integrals, Series, and Products (Academic Press, New York,
1980), 7th edition, edited by A. Jeffrey and D. Zwillinger, Sec. 6.5-6.7.



\bibitem{sherafati3} M. Sherafati and S. Satpathy, AIP Conf. Proc. \textbf{1461}, 24 (2012).


\end{thebibliography}
\end{document}